\RequirePackage{lineno}
\documentclass[twocolumn,showpacs,preprintnumbers,amsmath,amssymb,prl,superscriptaddress]{revtex4}
\usepackage{graphicx}
\usepackage{dcolumn}
\usepackage{bm}
\modulolinenumbers[1]
\newenvironment{color}[3]
{
}{
}
\newcommand{\grey}[1]     {}

\begin{document}
\title{Higher Moments of Net-proton Multiplicity Distributions at RHIC}
\medskip
\affiliation{Argonne National Laboratory, Argonne, Illinois 60439, USA}
\affiliation{University of Birmingham, Birmingham, United Kingdom}
\affiliation{Brookhaven National Laboratory, Upton, New York 11973, USA}
\affiliation{University of California, Berkeley, California 94720, USA}
\affiliation{University of California, Davis, California 95616, USA}
\affiliation{University of California, Los Angeles, California 90095, USA}
\affiliation{Universidade Estadual de Campinas, Sao Paulo, Brazil}
\affiliation{University of Illinois at Chicago, Chicago, Illinois 60607, USA}
\affiliation{Creighton University, Omaha, Nebraska 68178, USA}
\affiliation{Czech Technical University in Prague, FNSPE, Prague, 115 19, Czech Republic}
\affiliation{Nuclear Physics Institute AS CR, 250 68 \v{R}e\v{z}/Prague, Czech Republic}
\affiliation{University of Frankfurt, Frankfurt, Germany}
\affiliation{Institute of Physics, Bhubaneswar 751005, India}
\affiliation{Indian Institute of Technology, Mumbai, India}
\affiliation{Indiana University, Bloomington, Indiana 47408, USA}
\affiliation{Alikhanov Institute for Theoretical and Experimental Physics, Moscow, Russia}
\affiliation{University of Jammu, Jammu 180001, India}
\affiliation{Joint Institute for Nuclear Research, Dubna, 141 980, Russia}
\affiliation{Kent State University, Kent, Ohio 44242, USA}
\affiliation{University of Kentucky, Lexington, Kentucky, 40506-0055, USA}
\affiliation{Institute of Modern Physics, Lanzhou, China}
\affiliation{Lawrence Berkeley National Laboratory, Berkeley, California 94720, USA}
\affiliation{Massachusetts Institute of Technology, Cambridge, MA 02139-4307, USA}
\affiliation{Max-Planck-Institut f\"ur Physik, Munich, Germany}
\affiliation{Michigan State University, East Lansing, Michigan 48824, USA}
\affiliation{Moscow Engineering Physics Institute, Moscow Russia}
\affiliation{City College of New York, New York City, New York 10031, USA}
\affiliation{NIKHEF and Utrecht University, Amsterdam, The Netherlands}
\affiliation{Ohio State University, Columbus, Ohio 43210, USA}
\affiliation{Old Dominion University, Norfolk, VA, 23529, USA}
\affiliation{Panjab University, Chandigarh 160014, India}
\affiliation{Pennsylvania State University, University Park, Pennsylvania 16802, USA}
\affiliation{Institute of High Energy Physics, Protvino, Russia}
\affiliation{Purdue University, West Lafayette, Indiana 47907, USA}
\affiliation{Pusan National University, Pusan, Republic of Korea}
\affiliation{University of Rajasthan, Jaipur 302004, India}
\affiliation{Rice University, Houston, Texas 77251, USA}
\affiliation{Universidade de Sao Paulo, Sao Paulo, Brazil}
\affiliation{University of Science \& Technology of China, Hefei 230026, China}
\affiliation{Shandong University, Jinan, Shandong 250100, China}
\affiliation{Shanghai Institute of Applied Physics, Shanghai 201800, China}
\affiliation{SUBATECH, Nantes, France}
\affiliation{Texas A\&M University, College Station, Texas 77843, USA}
\affiliation{University of Texas, Austin, Texas 78712, USA}
\affiliation{Tsinghua University, Beijing 100084, China}
\affiliation{United States Naval Academy, Annapolis, MD 21402, USA}
\affiliation{Valparaiso University, Valparaiso, Indiana 46383, USA}
\affiliation{Variable Energy Cyclotron Centre, Kolkata 700064, India}
\affiliation{Warsaw University of Technology, Warsaw, Poland}
\affiliation{University of Washington, Seattle, Washington 98195, USA}
\affiliation{Wayne State University, Detroit, Michigan 48201, USA}
\affiliation{Institute of Particle Physics, CCNU (HZNU), Wuhan 430079, China}
\affiliation{Yale University, New Haven, Connecticut 06520, USA}
\affiliation{University of Zagreb, Zagreb, HR-10002, Croatia}

\author{M.~M.~Aggarwal}\affiliation{Panjab University, Chandigarh 160014, India}
\author{Z.~Ahammed}\affiliation{Lawrence Berkeley National Laboratory, Berkeley, California 94720, USA}
\author{A.~V.~Alakhverdyants}\affiliation{Joint Institute for Nuclear Research, Dubna, 141 980, Russia}
\author{I.~Alekseev~~}\affiliation{Alikhanov Institute for Theoretical and Experimental Physics, Moscow, Russia}
\author{J.~Alford}\affiliation{Kent State University, Kent, Ohio 44242, USA}
\author{B.~D.~Anderson}\affiliation{Kent State University, Kent, Ohio 44242, USA}
\author{D.~Arkhipkin}\affiliation{Brookhaven National Laboratory, Upton, New York 11973, USA}
\author{G.~S.~Averichev}\affiliation{Joint Institute for Nuclear Research, Dubna, 141 980, Russia}
\author{J.~Balewski}\affiliation{Massachusetts Institute of Technology, Cambridge, MA 02139-4307, USA}
\author{L.~S.~Barnby}\affiliation{University of Birmingham, Birmingham, United Kingdom}
% ...

\collaboration{STAR Collaboration}\noaffiliation

\date{\today}
\begin{abstract}
We report the first measurements of the kurtosis ($\kappa$), skewness ($\it {S}$) and variance
($\sigma^2$) of net-proton multiplicity ($N_{\mathrm p} - N_{\bar{\mathrm p}}$) 
distributions at midrapidity 
for Au+Au collisions at $\sqrt{s_{\mathrm {NN}}}$ = 19.6, 62.4, and 200 GeV 
corresponding to baryon chemical potentials ($\mu_{B}$) between 200 - 20 MeV. 
Our measurements of the products $\kappa$$\sigma^2$ and $\it{S}$$\sigma$,
which can be related to theoretical calculations sensitive to baryon number 
susceptibilities and long range correlations, 
are constant as functions of collision centrality.
We compare these products with results from lattice QCD and various models without a 
critical point and study the $\sqrt{s_{\mathrm {NN}}}$ dependence of $\kappa$$\sigma^2$.
From the measurements at the three beam energies, we find no evidence for a 
critical point in the QCD phase diagram for $\mu_{B}$ below 200 MeV.
\end{abstract}
\pacs{25.75.Gz,12.38.Mh,21.65.Qr,25.75.-q,25.75.Nq}
\maketitle
One of the major goals of the heavy-ion collision program is to explore 
the QCD phase diagram~\cite{starwhitepaper}. Finite temperature lattice 
QCD calculations~\cite{crossover} at baryon chemical potential $\mu_{\mathrm B}$ = 0 suggest a cross-over 
above a critical temperature ($T_{\mathrm c}$) $\sim$  170 -- 190 MeV~\cite{transitiontemp} 
from a system with hadronic degrees of freedom to a system where the 
relevant degrees of freedom are quarks and gluons. Several QCD based 
calculations (see e.g~\cite{firstorder}) find 
the quark-hadron phase transition 
to be first order at large $\mu_{\mathrm B}$. The point in the QCD phase 
plane ($T$ vs. $\mu_{\mathrm B}$) where the first order phase transition ends is 
the QCD Critical Point (CP)~\cite{qcp,qcp1}. Attempts are being made to 
locate the CP both experimentally and theoretically~\cite{bmqm09}. Current
theoretical calculations are highly uncertain about location of the CP. 
Lattice QCD calculations at finite  $\mu_{\mathrm B}$  
face numerical challenges in computing. The experimental 
plan is to vary the center of mass energy ($\sqrt{s_{\mathrm {NN}}}$) of 
heavy-ion collisions to scan the phase plane~\cite{bes} and at each energy, 
search for signatures of the CP that could survive the 
time evolution of the system~\cite{survival}.

In a static, infinite medium, the correlation length ($\xi$) diverges 
at the CP. $\xi$ is related to various moments of the distributions of 
conserved quantities such as net-baryons, net-charge, and net-strangeness~\cite{volker}.
Typically variances ($\sigma^2$ $\equiv$ $\left\langle (\Delta N)^2 \right\rangle$; 
$\Delta N = N - M$; $M$ is the mean) of these distributions are related to $\xi$ as 
$\sigma^2$ $\sim$ $\xi^2$~\cite{stephanovprd}. 
Finite size and time effects in heavy-ion collisions put constraints on the 
values of $\xi$. A theoretical calculation suggests $\xi$ $\approx$ 2-3 fm 
for heavy-ion collisions~\cite{krishnaxi}. It was recently shown that higher 
moments of distributions of conserved quantities, measuring deviations from a 
Gaussian, have a sensitivity to CP fluctuations that is better than that of $\sigma^2$, 
due to a stronger dependence on $\xi$~\cite{stephanovmom}. 
The numerators in skewness ($\it {S}$ = $\left\langle (\Delta N)^3 \right\rangle/\sigma^{3}$) goes 
as $\xi^{4.5}$ and 
kurtosis ($\kappa$ = [$\left\langle (\Delta N)^4 \right\rangle/\sigma^{4}$] - 3) 
goes as $\xi^7$. A crossing of the phase boundary can manifest itself by 
a change of
sign of $\it{S}$ as a function of energy density~\cite{stephanovmom,asakawa}.

Lattice calculations and QCD-based models show that moments of 
net-baryon distributions are related to baryon number ($\Delta N_{\mathrm B}$)
susceptibilities ($\chi_{\mathrm B} = \frac{\left\langle (\Delta N_{\mathrm B})^{2}\right\rangle}{VT}$;
{\it V} is the volume)~\cite{latticesus}. 
The product $\kappa$$\sigma^2$, related to the ratio of fourth order ($\chi^{(4)}_{\mathrm B}$) 
to second order ($\chi^{(2)}_{\mathrm B}$) susceptibilities, shows a large deviation from unity 
near the CP~\cite{latticesus}. Experimentally measuring event-by-event net-baryon numbers 
is difficult. However, the net-proton multiplicity 
($N_{\mathrm p} - N_{\bar{\mathrm p}}$ = $\Delta N_{\mathrm p}$) 
distribution is measurable. Theoretical calculations have shown 
that $\Delta N_{\mathrm p}$ 
fluctuations reflect 
the singularity of the charge and baryon number susceptibility as expected at 
the CP~\cite{hatta}. Non-CP model calculations (discussed later in the paper)
show that the inclusion of other baryons does not add to the sensitivity of 
the observable. This letter reports the first measurement of higher moments 
of the $\Delta N_{\mathrm p}$ distributions from Au+Au collisions to search 
for signatures of the CP.

\begin{figure}[htp]
\includegraphics[scale=0.4]{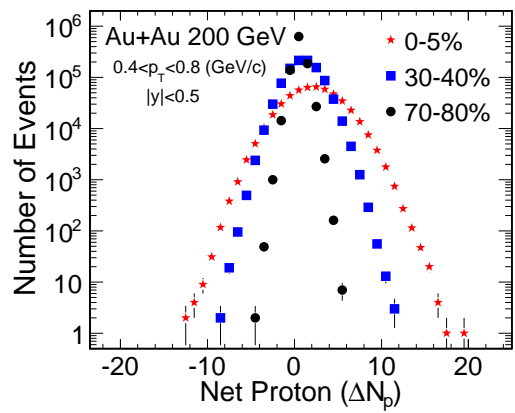}
\caption
{(Color online) 
$\Delta N_{\mathrm p}$ multiplicity distribution in Au+Au collisions at 
$\sqrt{s_{\mathrm {NN}}}$ = 200 GeV for various collision centralities 
at midrapidity ($\mid y \mid < 0.5$). The statistical errors are shown. 
}
\label{Fig1}
\end{figure}
The data presented in the paper are obtained using the Time Projection
Chamber (TPC) of the Solenoidal Tracker at RHIC (STAR)~\cite{star}. 
The event-by-event proton ($N_{\mathrm p}$) and anti-proton ($N_{\bar{\mathrm p}}$) multiplicities
are measured for Au+Au minimum bias events at $\sqrt{s_{\mathrm {NN}}}$ = 19.6, 62.4, 
and 200 GeV for collisions occurring within 30 cm of 
the TPC center along the beam line. 
The numbers of events analyzed are 4$\times 10^{4}$, 5$\times 10^{6}$,
and 8$\times 10^{6}$ for  $\sqrt{s_{\mathrm {NN}}}$ = 19.6, 62.4, and 200 GeV, 
respectively. Centrality selection  utilized the uncorrected charged particle 
multiplicity within pseudorapidity $\mid \eta \mid$ $<$ 0.5, measured by the TPC. 
For each centrality, the average numbers of participants 
($\langle N_{\mathrm {part}} \rangle$) are 
obtained by Glauber model calculations. The $\Delta N_{\mathrm p}$ measurements are 
carried out at midrapidity ($\mid y \mid$ $<$ 0.5) in the range 
0.4 $<$ $p_{\mathrm T}$ $<$ 0.8 GeV/$c$. Ionization energy loss ($dE/dx$) of charged 
particles in the TPC was used to identify the inclusive $p$($\bar{p}$)~\cite{starprc}. 
To suppress the contamination from secondary protons, we required each 
$p(\bar{p})$ track to have a minimum $p_{\mathrm T}$ of 0.4 GeV/c 
and a distance of closest approach (DCA) to the primary vertex
of less than 1 cm~\cite{starprc}. The $p_{\mathrm T}$ range used includes 
approximately 35-40\% of the total $p+ \bar{p}$ 
multiplicity at midrapidity. 
$\Delta N_{\mathrm p}$ was not corrected for reconstruction efficiency. 
Typical $\Delta N_{\mathrm p}$ distributions from 70-80\%, 30-40\%, and 0-5\% 
Au+Au collision centralities are shown in Fig.~\ref{Fig1}. 

\begin{figure}[htp]
\includegraphics[width=0.54\textwidth]{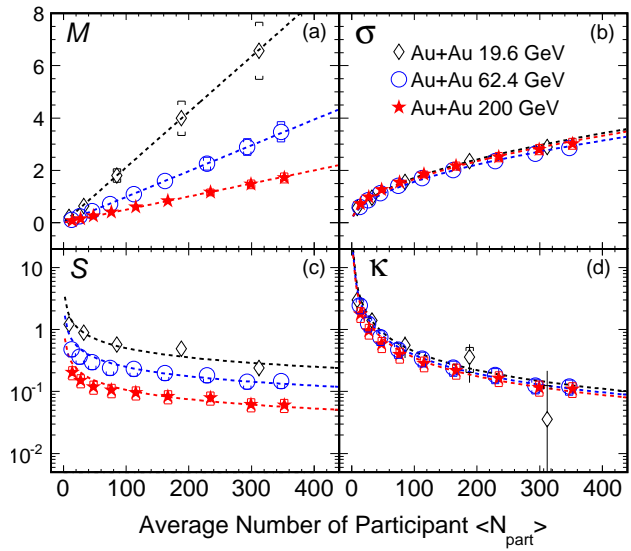}
\caption
{(Color online) Centrality dependence of moments of $\Delta N_{\mathrm p}$ distributions
for Au+Au collisions at $\sqrt{s_{\mathrm {NN}}}$ = 19.6, 62.4, and 200 GeV. 
The lines are the expected values from the central limit theorem. 
Error bars are statistical and caps are systematic errors.
}
\label{Fig2}
\end{figure}
The four moments ($M$, $\sigma$, $\it{S}$, and $\kappa$) which describe the shape
of the $\Delta N_{\mathrm p}$ distributions at various collision
energies are plotted as a function of $\langle N_{\mathrm {part}} \rangle$
in Fig.~\ref{Fig2}. 
The typical statistical errors on $\sigma$, $\it{S}$, 
and $\kappa$ for central Au+Au collisions at 200 GeV are 0.2\%, 11\% and 16\% 
respectively. The $M$ shows a linear variation with $\langle N_{\mathrm {part}}\rangle$
and increases as $\sqrt{s_{\mathrm {NN}}}$ decreases, in accordance with the energy
and centrality dependence of baryon transport~\cite{bes}. 
The variation of $M$ within a centrality bin has been taken into account in higher moment
calculations. The $\sigma$ increases with $\langle N_{\mathrm {part}} \rangle$. 
The values are similar for three beam energies studied. 
The $\it{S}$ is positive and decreases as $\langle N_{\mathrm {part}}\rangle$ increases 
for a given collision energy. The values also decrease as $\sqrt{s_{\mathrm {NN}}}$
increases. This indicates that the distributions become symmetric for 
more central collisions 
and for higher beam energies. The $\kappa$ decreases as $\langle N_{\mathrm {part}}\rangle$ 
increases, but is similar for all three $\sqrt{s_{\mathrm {NN}}}$ studied. 

Experimentally it is difficult to correct such observables for the particle reconstruction 
efficiency on an event-by-event basis. Construction of observables independent of the efficiency, 
such as factorial moments, leads to loss of one-to-one correspondence 
with higher moments~\cite{fac}, and significant difficulty
in comparing to theoretical expectations. 
We have investigated the effects of the detector and track reconstruction efficiencies 
by comparing the moments of the $\Delta N_{\mathrm p}$ distribution using the events 
from a heavy-ion event generator model HIJING (ver.1.35)~\cite{hijing} 
and the moments of the reconstructed 
$\Delta N_{\mathrm p}$, after passing the same events through a realistic 
GEANT detector simulation.
The difference between the two cases for the $\sigma$, $\it{S}$ and $\kappa$ are about an order 
of magnitude smaller than their absolute values. Typical values of such differences for central Au+Au 200 GeV 
collisions are -0.37$\pm$0.05, 0.02$\pm$0.05 and -0.06$\pm$0.12 for 
 $\sigma$, $\it{S}$, and $\kappa$, respectively.
These results indicate that the effects on the shape of the distributions are small. 
The effect on the yields of $p$($\bar{p}$) is discussed elsewhere~\cite{starprc,bes}. 
The systematic errors are estimated by varying the following requirements 
for $p(\bar{p})$ tracks: 
DCA, track quality reflected by the number of fit points used in track reconstruction,
and the $dE/dx$ selection criteria for $p(\bar{p})$ 
identification. The typical systematic 
errors are of the order 10\% for $\it {M}$ and $\sigma$, 25\% on $\it{S}$  
and 30\% on $\kappa$. The statistical and systematic (caps) errors are presented 
separately in the figures.

To understand the evolution of centrality dependence of moments in Fig.~\ref{Fig2}, 
we invoke the central limit theorem (CLT) and consider the distribution at any given 
centrality $i$ to be a superposition of several independent source distributions.
We assume the average number of the sources for a given centrality to be equal to some 
number $C$ times the corresponding $\langle N_{\mathrm {part}}\rangle$, and obtain [21]: 
\begin{equation}
M_i = C M_{\mathrm x} \langle N_{\mathrm {part}}\rangle_i,
\end{equation}
\begin{equation}
\sigma^{2}_i = C \sigma_{\mathrm x}^{2} \langle N_{\mathrm {part}}\rangle_{i}, 
\end{equation}
\begin{equation}
\it{S}_{i} = \it{S}_{\mathrm x}/ [\sqrt{C\langle N_{\mathrm {part}}\rangle_{i}}],
\end{equation}
and 
\begin{equation}
\kappa_{i} = \kappa_{\mathrm x}/[C\langle N_{\mathrm {part}}\rangle_{i}].
\end{equation}
The various moments of the parent distribution $M_{\mathrm x}$, $\sigma_{\mathrm x}$, $\it{S}_{\mathrm x}$, $\kappa_{\mathrm x}$ and constant $C$ have been determined from fits to data. 
The dashed lines in Fig.~\ref{Fig2} show the expectations from the CLT. 
The $\chi^{2}/\rm{ndf}$ between the CLT expectations 
and data are $<$ 1.5 for all the moments presented. 
If collision centrality reflects the system volume, then the results in Fig.~\ref{Fig2}
which approximate baryon number susceptibilities suggest that the susceptibilities do
not change with the volume~\cite{crossover}. 
Deviations from $\langle N_{\mathrm {part}} \rangle$ scaling could
indicate new physics such as might result from the CP.

\begin{figure}[htp]
\includegraphics[scale=0.45]{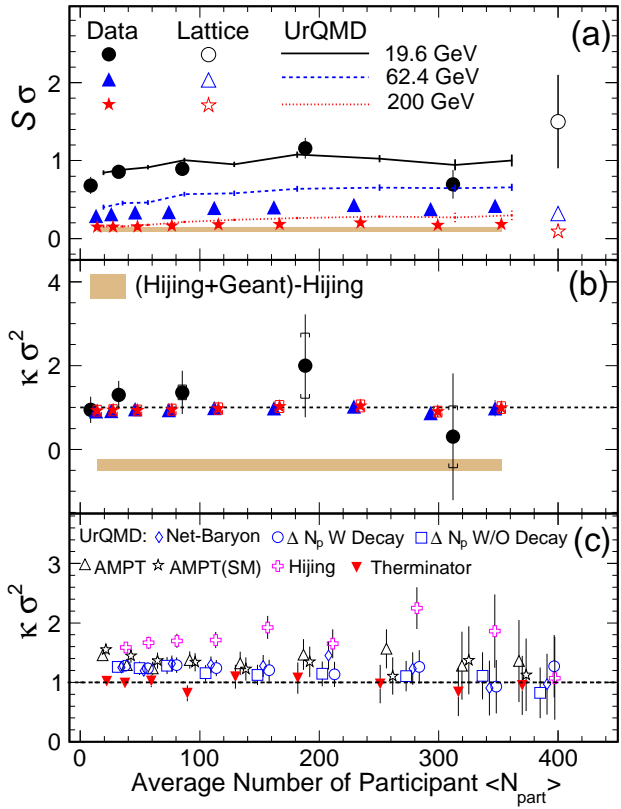}
\caption
{(Color online) Centrality dependence of (a) $\it{S}$$\sigma$
and (b) $\kappa$$\sigma^2$ for $\Delta N_{\mathrm p}$ in Au+Au collisions at 
$\sqrt{s_{\mathrm {NN}}}$ = 19.6, 62.4, and 200 GeV compared
to various model calculations. 
The shaded band for $\it{S}$$\sigma$ and $\kappa$$\sigma^2$ reflects contributions from 
the detector effects. (c) shows the model expectations for 
$\kappa$$\sigma^2$ from various physical effects in Au+Au collisions at 200 GeV. 
The lattice QCD results are for net-baryons corresponding 
to central collisions~\cite{qcp1}. See text for more details. 
}
\label{Fig3}
\end{figure}

To get a microscopic view, we present two observables, 
$\it{S}$$\sigma$ and $\kappa$$\sigma^2$,  which can be used to
search for the CP. 
These products will be constants as per the CLT and other likely non-CP scenarios, 
as seen from the dependences on $\langle N_{\mathrm {part}} \rangle$ discussed above. 
These observables are related to 
the ratio of baryon number susceptibilities ($\chi_{\mathrm B}$) at a given temperature ($T$) 
computed in QCD models as: $\it{S}$$\sigma$ = $\frac{\chi^{(3)}_{\mathrm B}/T}{\chi^{(2)}_{\mathrm B}/T^2}$ and
$\kappa$$\sigma^2$ = $\frac{\chi^{(4)}_{\mathrm B}}{\chi^{(2)}_{\mathrm B}/T^2}$~\cite{qcp1}. 
Close to the CP, models predict the net-baryon
number distributions to be non-Gaussian and susceptibilities to diverge causing $\it{S}$$\sigma$
and $\kappa$$\sigma^2$ to deviate from being constants and have large values. 
Figure~\ref{Fig3} shows that $\it{S}$$\sigma$ and $\kappa$$\sigma^2$ 
for Au+Au collisions at $\sqrt{s_{\mathrm {NN}}}$ = 19.6, 62.4, and 200 GeV are constants 
as a function of $\langle N_{\mathrm {part}} \rangle$. 

In Fig.~\ref{Fig3}(a), lattice QCD results on $\it{S}$$\sigma$ for net-baryons in 
central collisions are found to agree with the measurements. Near the CP, the
system will deviate from equilibrium~\cite{krishnaxi} and results from lattice QCD,
which assumes equilibrium, should not be consistent with the data. 
These lattice calculations, which predict a CP around 
$\mu_{\mathrm B}$ $\sim$ 300 MeV, are carried out using two-flavor QCD with number of lattice sites
in imaginary time to be 6 and mass of pion around 230 MeV~\cite{qcp1}.
The ratios of the non-linear susceptibilities at finite $\mu_{\mathrm B}$ are obtained
using Pad\'e approximant resummations of the quark number susceptibility series. 
The freeze-out parameters as a function of $\sqrt{s_{\mathrm {NN}}}$ are taken from~\cite{cleymans}
and $T_{\mathrm c}$ = 175 MeV. 

To understand the various non-CP 
physics background contribution 
to these observables, in Fig.~\ref{Fig3} we also present the results for 
the net-proton distribution as a function of $\langle N_{\mathrm {part}} \rangle$ 
from UrQMD (ver.2.3)~\cite{urqmd}, HIJING~\cite{hijing}, AMPT (ver.1.11)~\cite{ampt}, and 
Therminator (ver.1.0)~\cite{thermus} models. 
The measurements are consistent with results from 
various non-CP models studied. In Fig.~\ref{Fig3}(c),
several model calculations from Au+Au collisions at 200 GeV are presented to 
understand the effect of the following on our observable: with (W) and 
without (W/O) resonance decays, 
inclusion of all baryons 
(both studied using UrQMD), jet-production (HIJING), coalescence mechanism of particle
production (AMPT String Melting, ver.2.11), thermal particle production (Therminator), rescattering (UrQMD
and AMPT). All model calculations are done using default versions and 
with the same kinematic coverage as for data. 
The $\kappa$$\sigma^2$ (Fig.~\ref{Fig3}b) and $\it{S}$$\sigma$ (Fig.~\ref{Fig3}a) 
are found to be 
constant for all the cases as a function of $\langle N_{\mathrm {part}} \rangle$.
This constant value can act as a baseline for the CP search.
QCD model calculations with CP predict a non-monotonic dependence of these observables
with $\langle N_{\mathrm {part}} \rangle$ and $\sqrt{s_{\mathrm {NN}}}$~\cite{stephanovmom}.

\begin{figure}[htp]
\includegraphics[width=0.52\textwidth]{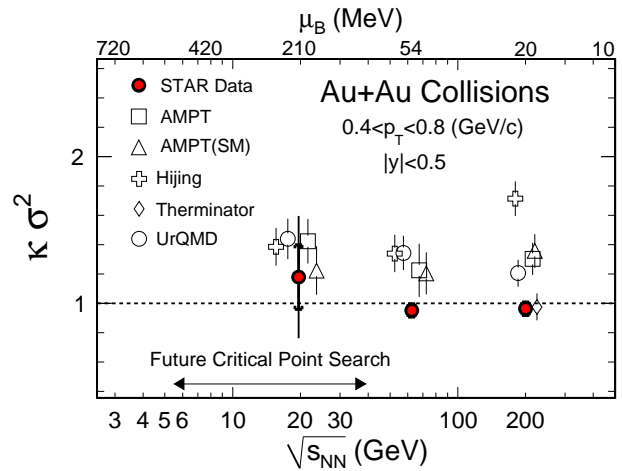}
\caption
{(Color online) $\sqrt{s_{\mathrm {NN}}}$ dependence of $\kappa$$\sigma^2$ for
net-proton distributions measured at RHIC. The results are compared to non-CP
model calculations (slightly shifted in $\sqrt{s_{\mathrm {NN}}}$). 
The left-right arrow at the bottom indicates the energy range for the CP search at RHIC.
}
\label{Fig4}
\end{figure}

Figure~\ref{Fig4} shows the energy dependence of $\kappa$$\sigma^2$ for $\Delta N_{\mathrm p}$, 
compared to several model calculations that do not include a CP. The experimental values plotted 
are average values for the centrality range studied; they 
are found to be consistent with unity. 
Also shown at the top of Fig.~\ref{Fig4} are the $\mu_{\mathrm B}$ values corresponding to 
the various $\sqrt{s_{\mathrm {NN}}}$~\cite{starprc,cleymans}. 
We observe no non-monotonic dependence with $\sqrt{s_{\mathrm {NN}}}$.
The results from non-CP models are constants as a function of $\sqrt{s_{\mathrm {NN}}}$ and 
have values between 1-2. The result from the thermal model is exactly unity. 
Within the ambit of the models studied, the observable changes little
with change in non-CP physics (such as due to change in $\mu_{\mathrm B}$, 
collective expansion and particle production) at the various energies studied. 
From comparisons to models and the lack of non-monotonic
dependence of $\kappa$$\sigma^2$ on $\sqrt{s_{\mathrm {NN}}}$ studied, we 
conclude that there is no indication from our measurements for a CP
in the region of the phase plane with  $\mu_{\mathrm B}$ $<$ 200 MeV.
It is difficult to rule out the existence of CP for the entire 
$\mu_{\mathrm B}$ region below 200 MeV. 
The extent to which these results can do that is guided by the following theoretical work.
One QCD based model including a CP ($\xi$ = 3 fm)
predicts the value of $\kappa$$\sigma^2$ to be at least a factor of 2 higher 
than the measurements presented ($\kappa$$\sigma^2$ $\sim$ 2.5, 35, 3700 for 
the CP at $\sqrt{s_{\mathrm {NN}}}$ =  200, 62.4, and 19.6 GeV, respectively)~\cite{stephanovmom}.  
In addition, the expectation of the extent of the critical region 
in $\mu_{\mathrm B}$ is thought to be about 100 MeV~\cite{qcp1,qcpext}.

In summary, the first measurements of the higher moments of the net-proton distributions 
at midrapidity ($\mid y\mid$$<$ 0.5) within 0.4 $<$ $p_{\mathrm T}$ $<$ 0.8 GeV/$c$ 
in Au+Au collisions at  
$\sqrt{s_{\mathrm {NN}}}$ = 19.6, 62.4, and 200 GeV have been presented. New observables
$\it{S}$$\sigma$ and $\kappa$$\sigma^2$ 
derived from the $\Delta N_{\mathrm p}$ distribution to search for the CP in heavy-ion collisions 
are discussed. These observables are found to be constant as a function of 
$\langle N_{\mathrm {part}} \rangle$ for all collisions
energies studied. This is consistent with expectations from the central limit theorem and
in general agreement with results from various models without the CP. 
The measured $\it{S}$$\sigma$ in central collisions are consistent with lattice QCD 
calculations of the ratio of third order to second order baryon number susceptibilities.
Within the uncertainties, $\kappa$$\sigma^2$ is found to be constant as a function of 
$\sqrt{s_{\mathrm {NN}}}$ studied. This trend is consistent with models without a CP
and in sharp contrast to models~\cite{stephanovmom} which include a CP
in this $\mu_{\mathrm B}$ range.
Our measurements show no evidence for a CP to be located 
at $\mu_{\mathrm B}$ values ${}^<_\sim$ 200 MeV in the QCD phase plane.
The RHIC beam energy (100 $<$ $\mu_{B}$ $<$ 550 MeV) 
scan will look for non-monotonic variation of $\kappa$$\sigma^2$ for net-protons as 
a function of $\sqrt{s_{\mathrm {NN}}}$ to locate the CP.

\indent We thank S. Gupta, F. Karsch, K. Rajagopal, K. Redlich and M. Stephanov for discussions.
We thank the RHIC Operations Group and RCF at BNL, and the NERSC Center 
at LBNL and the Open Science Grid consortium 
for their support. This work was supported in part by the Offices of NP 
and HEP in the U.S. DOE Office of Science, the U.S. NSF, the Sloan 
Foundation, the DFG cluster of excellence `Origin and Structure of 
the Universe' of Germany, CNRS/IN2P3, RA, RPL, and EMN of France, STFC and EPSRC 
of the United Kingdom, FAPESP of Brazil, the Russian Ministry of Sci. 
and Tech., the NNSFC, CAS, MoST, and MoE of China, IRP and GA of the 
Czech Republic, FOM of the Netherlands, DAE, DST, and CSIR of the 
Government of India, the Polish State Committee for Scientific 
Research,  and the Korea Sci. $\&$ Eng. Foundation.

\end{document}